\documentclass[letterpaper,fp]{jpsj3}
\usepackage{txfonts}
\usepackage{graphicx,color}
\usepackage{dcolumn}
\usepackage{bm}
\usepackage{amsmath}
\usepackage{amsfonts}
\usepackage{amssymb}
\usepackage{subfigure}

\begin{document}

\title{
Acoustic wave propagation through a supercooled liquid: A normal mode analysis
}

\author{Yuki Matsuoka\thanks{Current address: Production Engineering Research Laboratory, Sumitomo Bakelite Co., Ltd., Shizuoka 426-0041, Japan},
Hideyuki Mizuno\thanks{Current address: Laboratory for Interdisciplinary Physics, UMR 5588, Universit\'e Grenoble 1 and CNRS, Saint Martin d'H\`eres, 38402, France},
Ryoichi Yamamoto\thanks{Email: ryoichi@cheme.kyoto-u.ac.jp}}
\inst{Department of Chemical Engineering, Kyoto University, Kyoto 615-8510, Japan}

\abst{
The mechanism of acoustic wave propagation in supercooled liquids is not yet fully understood since the vibrational dynamics of supercooled liquids are strongly affected by their amorphous inherent structures. In this paper, the acoustic wave propagation in a supercooled model liquid is studied by using normal mode analysis. Due to the highly disordered inherent structure, a single acoustic wave is decomposed into many normal modes in broad frequency range. This causes the rapid decay of the acoustic wave and results in anomalous wavenumber dependency of the dispersion relation and the rate of attenuation.
}

\kword{supercooled liquid, disordered structure, vibrational dynamics, acoustic wave propagation, normal mode analysis}

\maketitle

\section{Introduction}
As liquids are cooled toward the glass transition temperature $T_g$, their mechanical properties become much different from those of ``normal liquids''. Such the supercooled liquids not only become more viscous but also become more elastic toward $T_g$ \cite{rheology}. Associated with drastic change of the mechanical properties, the acoustic properties also differ notably from those of normal liquids. In fact, due to marked elastic properties, the supercooled liquids support the transverse acoustic waves for a long distance like solids \cite{mountain_1995,ahluwalia_1998}. Although the acoustic wave propagation is a fundamental issue in material science, it is not still fully understood for the supercooled liquids \cite{hiwatari_1990,roux_1989}.

In the supercooled liquids, dynamics of molecules are strongly influenced by the ``inherent structure'', i.e., the underlying potential energy landscape characteristic of their amorphous structures \cite{sastry_1998,grigera_2002}. Relaxation processes can be well described by two characteristic dynamics, {\it i.e.} the vibrational dynamics within local potential minima and the activation dynamics between them whose dynamics is characterized by the so-called $\alpha$-relaxation time $\tau_\alpha$ \cite{cavagna_2009}. As two time scales of these two dynamics are well separated by several orders of magnitude in the supercooled state, we can examine the acoustic wave propagation by analyzing the vibrational dynamics of molecules around local minima of the inherent structure.

In order to investigate the vibrational dynamics, ``normal mode analysis'' is used as a powerful tool \cite{mazzacurati_1996,ruocco_2000,angelani_2000,grigera_2003,schober_2004}, where the vibrational dynamics are decomposed into many harmonic oscillations. In the case of perfect crystals which have periodic and ordered structures, a single acoustic wave mode is described by a single normal mode, which is well known as ``phonon". On the other hand, in the case of amorphous materials including supercooled liquids, due to their disordered structures, a single propagating plane wave cannot be described by a single normal mode, but a superposition of several different normal modes with different frequencies, where the dispersion relation is determined from the peak frequency and the attenuation rate is determined from the half-width of the spectrum density function of the normal mode frequencies. Although the same information can be obtained via the dynamic structure factor, which can be measured by scattering experiments \cite{shapiro_1966,sette_1998,bove_2005} and calculated also often in numerical studies \cite{rahman_1976,grest_1982,hiwatari_1990,horbach_2001}, the normal mode analysis enables us to connect properties of propagating acoustic waves with the normal modes of the amorphous inherent structures. This is an apparent advantage of the normal mode analysis.

In the present study, we examined the acoustic wave propagation through a supercooled model liquid by means of the normal mode analysis on the inherent structure. By decomposing a single acoustic wave into many normal modes with different frequencies, we obtained a clear picture on the mechanism of the acoustic wave propagation and attenuation which are highly affected by the disordered structure.

\section{Simulation model}
We used $3$-dimensional model liquid which is a equimolar binary mixture composed of species $1$ (small) and species $2$ (large). The particles interact via the soft-sphere potential,
\begin{equation}
\phi_{a b}(r)= \epsilon (\sigma_{a b}/r)^{12}, \quad \sigma_{a b} = (\sigma_a + \sigma_b)/2, \label{softcore}
\end{equation}
where $r$ is the distance between two particles, $\sigma_a$ is the particle diameter, and $a,b \in 1,2$. The cut-off radius for the potential was set at $3\sigma_{ab}$. The number of the particles is $N=N_1+N_2=5,000$ in a cube of constant volume $V$ under periodic boundary condition. The particle density was fixed at the relatively high value of $\rho=(N_1+N_2)/V=0.8 \sigma_1^{-3}$, where the system length was $L=V^{1/3}=18.4 \sigma_1$. The mass ratio was $m_2/m_1=2.0$, and the diameter ratio was $\sigma_2/\sigma_1=1.2$. This diameter ratio avoided crystallization of the system and ensured that an amorphous supercooled state formed at low temperatures \cite{miyagawa_1991}. We first equilibrated the system in a supercooled state at the temperature $T=0.289 \epsilon/k_B$ by using the $NVT$ molecular-dynamics (MD) simulation. Note that the melting temperature $T_m$ of this system was reported to be around $0.772 \epsilon/k_B$ \cite{miyagawa_1991}. Then, we obtained the inherent structure of the equilibrated system by means of the steepest descent method to determine the local minimum of the potential energy landscape. In the following, $\sigma_1$, $\epsilon/k_B$, and $\tau=(m_1\sigma_1^2/\epsilon)^{1/2}$ are used as units of length, temperature, and time, respectively.

\section{Normal mode analysis}
In the present study, we performed the normal mode analysis on the inherent structure. The dynamical matrix $\bm{V}$, the elements of which are the second derivatives of the potential energy $\Phi(\bm{r}_i)=\sum_{i=1}^N\sum_{j>i}\phi_{ij}(r_{ij})$ in mass-weighted generalized coordinates $\sqrt{m_i} \bm{r}_i$, was calculated. The eigenvalues and eigenvectors of $\bm{V}$ were then determined. If the number of particles is $N$, then $\bm{V}$ is a $3N \times 3N$ Hessian matrix, and its elements $\bm{V}_{ij}$ is a $3 \times 3$ matrix whose components are derivatives of the potential energy with respect to coordinates of particles $i$ and $j$ defined by
\begin{equation}
\bm{V}_{ij}=\frac{1}{\sqrt{m_i}\sqrt{m_j}}\left( \frac{\partial^2 \phi_{ij}(r_{ij})}{\partial \bm{r}_i \partial \bm{r}_j}\right)_0, \quad (i,j=1,2,\dots,N),
\end{equation}
where $(\ldots)_0$ indicates that the values were calculated for the inherent structure. We emphasize that the dynamic matrix $\bm{V}$ can be derived only from information regarding the inherent structure, i.e., the static particle configuration. Once this matrix had been simplified to a diagonal matrix, the $3N$ eigenvalues $\lambda_k \ (k=1,2,\dots,3N)$ and the $3N$ mutually perpendicular $3N$-dimensional eigenvectors $\bm{a}^k$ corresponding to the values of $\lambda_k$ were obtained. Here we normalized each eigenvector $\bm{a}^k$ by $\bm{a}^k \cdot \bm{a}^k = 1$. The eigen frequencies of the matrix were then determined from the relationship $\omega_k=\sqrt{\lambda_k}$. Because the above analysis employs second derivatives of the potential, it is essential to avoid truncation error in the cut-off radius of the potential. Therefore, a modification was applied to the interactions between particles in Eq. (\ref{softcore}) to prevent discontinuities in the interparticle potential energy and the force at the cut-off radius in the process of minimizing the energy and the normal mode analysis \cite{comsim}.

\begin{figure}[t]
\begin{center}
\includegraphics*[scale=0.55]{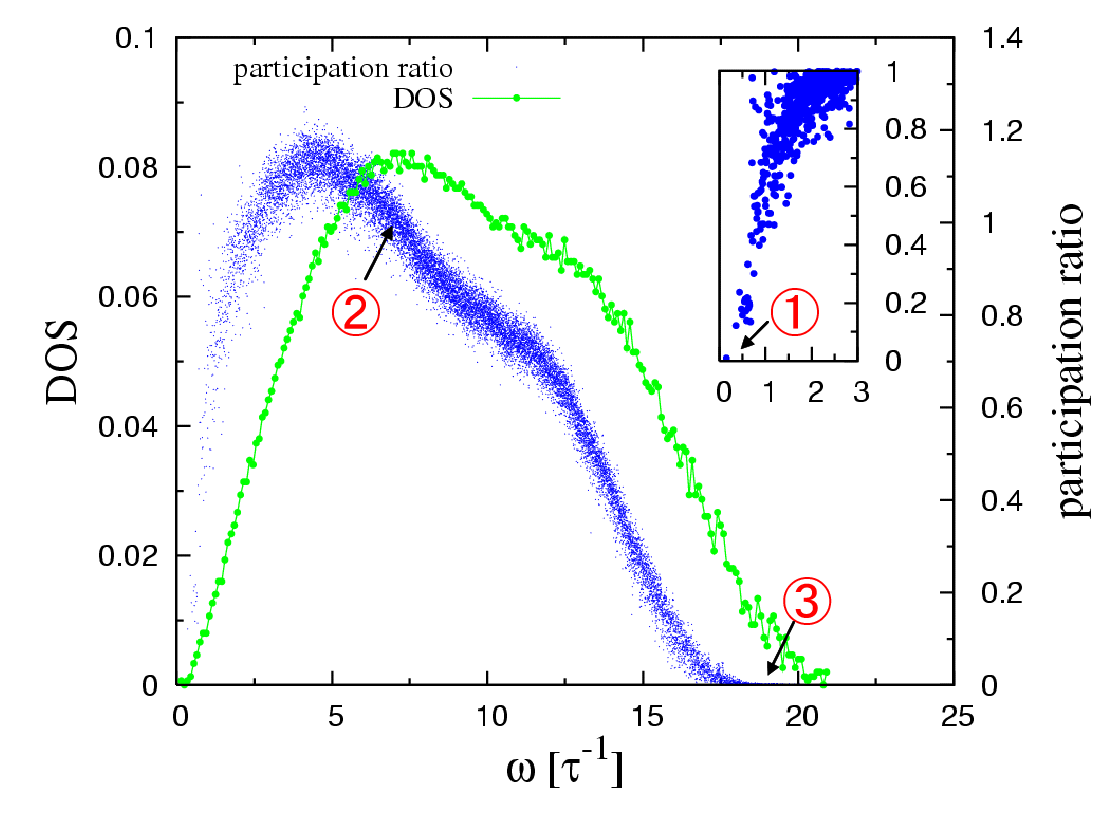}
\end{center}
\vspace{-.7cm}
\caption{
The density of states (DOS) and the participation ratio $p_k$. Green lines represent DOS frequency distributions, and blue dots represent participation ratios of modes. The inset on the upper right is a close-up of the participation ratio in the low frequency region of the spectrum. The eigenvectors of modes labeled with numbers are plotted in Fig. \ref{map_eigenvec}.
}
\label{dos_part}
\end{figure}

The vibrational frequency $\omega$ found by the normal mode analysis is classified into one of several different ranges, and the number of vibrational modes in each range is calculated using the density of states (DOS) distribution function
\begin{equation}
{\rm{DOS}}(\omega)=\frac{1}{3N}\sum_{k=1}^{3N}\delta(\omega-\omega_k),
\end{equation}
where $\delta$ is the Kronecker delta. Figure \ref{dos_part} shows the frequency distribution of DOS (green lines). Our result agrees well with that derived in Ref. \citenum{grigera_2003} (Fig. 2(c)), where the DOS was numerically calculated for the similar system as the present one.

\begin{figure}[t]
\begin{center}
\includegraphics*[scale=0.40, angle=270]{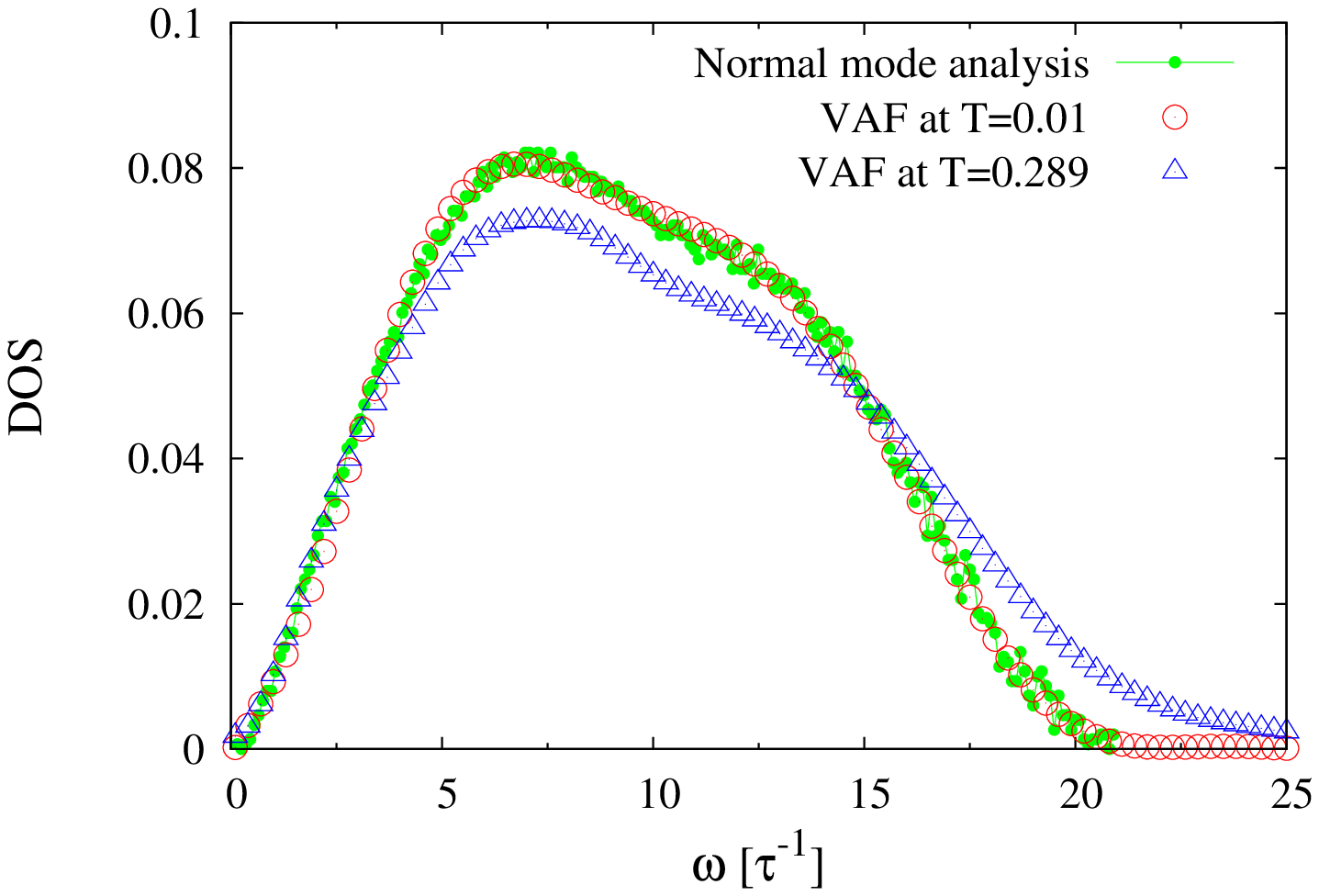}
\end{center}
\vspace{-.3cm}
\caption{
The comparison of the density of states (DOS) between the normal mode analysis and the Fourier transformation of the velocity autocorrelation function (VAF).
The same DOS of the normal mode analysis as plotted in Fig. \ref{dos_part} is shown.
The VAF is obtained at two states: the supercooled state ($T=0.289$) and the glass state ($T=0.01$).
}
\label{dos_com}
\end{figure}

The DOS can be also obtained from the Fourier transformation of the velocity autocorrelation function (VAF) \cite{shintani_2008}.
In the present study, we also calculated the DOS from the VAF to confirm the validity of our normal mode analysis.
We performed the $NVE$ MD simulations at two states: the supercooled state ($T=0.289$) and the glass state ($T=0.01$).
Note that the glass state was obtained by cooling the system from $T=0.289$ to $T=0.01$ with a quench rate of $dT/dt=3 \times 10^{-5}$.
After cooling, we relaxed the quenched glass sample for $t=10^5$ in the $NVT$ ensemble.
The Fourier transformation of the VAF was then calculated from the $NVE$ trajectory as
\begin{equation}
{\rm{DOS}}_{\text{VAF}}(\omega) = \int \frac{dt}{2 \pi} \exp (i \omega t) \frac{\left< \sum_{i=1}^N \sqrt{m_i}\bm{v}_i(t) \cdot \sqrt{m_i}\bm{v}_i(0) \right>}{\left< \sum_{i=1}^N \sqrt{m_i}\bm{v}_i(0) \cdot \sqrt{m_i}\bm{v}_i(0) \right>},
\end{equation}
where $\bm{v}_i$ is the velocity of the particle $i$, and $\left< \right>$ means the ensemble average over the initial time $0$.
We compare the DOS between the normal mode analysis and the Fourier transformation of the VAF in Fig. \ref{dos_com}.
At the supercooled state $T=0.289$, the VAF exhibits a little different DOS from the normal mode analysis.
This difference is due to the non-linear vibrational motions of particles.
Although the non-linear vibrations can make some effects on the sound wave propagation, we checked that such the non-linear effects are very small in the supercooled state.
In this study, we examined the sound wave propagation in the framework of the harmonic oscillations, so we did not consider the non-linear vibrations.
On the other hand, the VAF coincides well with the normal mode analysis at the glass state $T=0.01$.
At $T=0.01$, the vibrational motions of particles can be completely described by the harmonic oscillators \cite{ruocco_2000}, and therefore the VAF catches the harmonic vibrations as does the normal mode analysis.
The validity of our calculation can be confirmed from the coincidence of the normal mode analysis and the VAF.

In Fig. \ref{dos_part}, we also show the participation ratio $p_k$ (blue dots), which is a quantification of the extent of participation by each mode and is expressed by the following equation:
\begin{equation}
p_k=\left\lbrace N\sum_{j=1}^N
\left(\frac{\bm{a}_j^k\cdot\bm{a}_j^k}{m_j}\right)^2\right\rbrace^{-1},
\label{part}
\end{equation}
where $k$ is the mode index, and $\bm{a}_j^k$ is the eigenvector of particle $j$ in mode $k$. Notice that $\bm{a}_j^k$ is the $3$-dimensional vector which is the particle $j$ component of $3N$-dimensional eigenvector $\bm{a}^k$. ``Localized mode'' means that only a minority of the particles have very large eigenvectors. If all particles have eigenvectors of identical magnitude, $p_k$ will be $1$; on the contrary, if one particle has a very large eigenvector and if the mode is spatially localized, $p_k$ will have the very low value of $1/N$.

\begin{figure*}[t]
\begin{center}
\subfigure[]{\includegraphics*[scale=0.6]{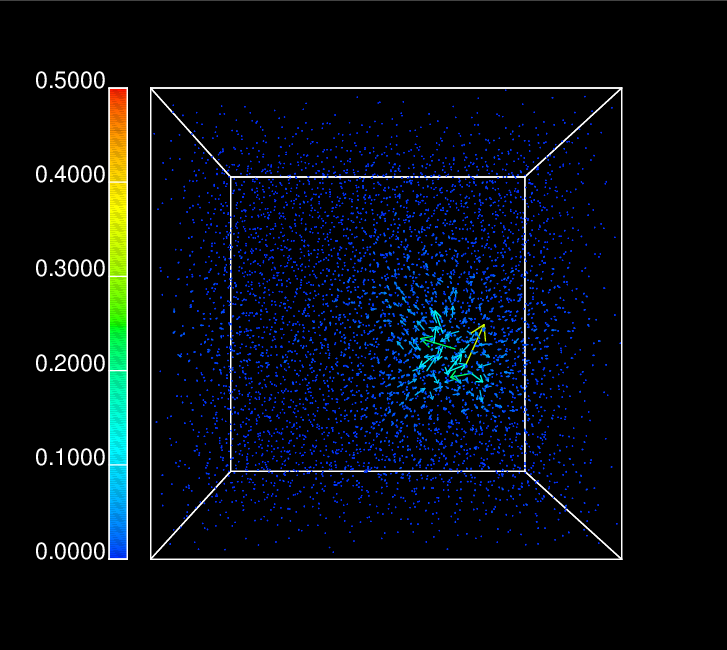}
\label{low_vec}}
\subfigure[]{\includegraphics*[scale=0.6]{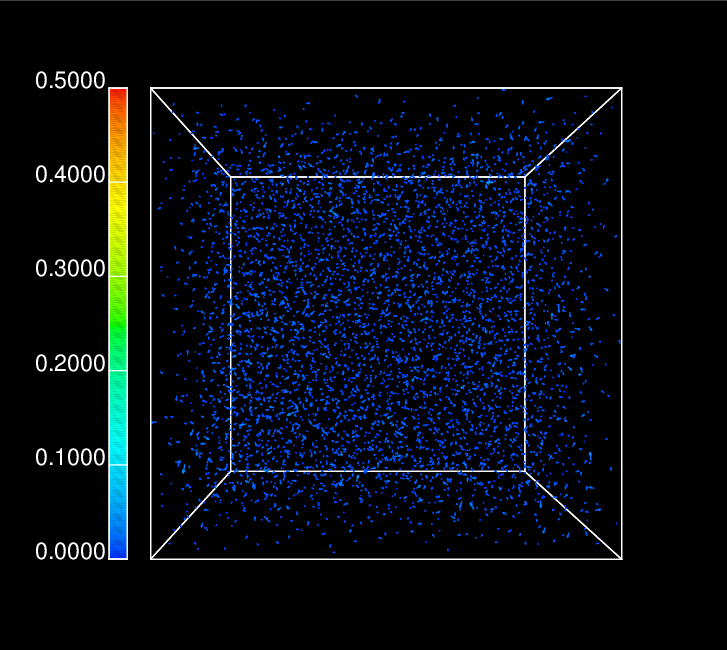}
\label{mid_vec}}
\subfigure[]{\includegraphics*[scale=0.6]{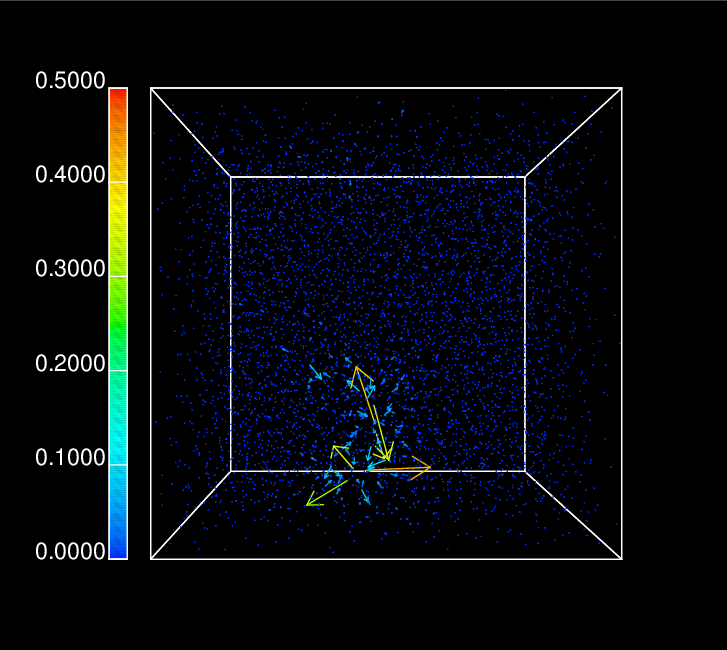}
\label{high_vec}}
\end{center}
\vspace{-0.5cm}
\caption{
Spatial distributions of eigenvectors at various frequencies.
(a) Low-frequency localized mode, where $\omega_k=0.154$ and $p_k=0.00946$.
(b) Medium-frequency delocalized mode at $\omega_k=7.000$ and $p_k=0.992$.
(c) High-frequency localized mode, where $\omega_k=18.99$ and $p_k=0.00120$.
Note that there are $N=5,000$ arrows (particles) in each box, and the linear length of the box is $L=18.4$. The origins of the vectors are located at the particle positions in the inherent structure (after energy minimization).
}
\label{map_eigenvec}
\end{figure*}

Figure \ref{dos_part} indicates very low participation in the lowest- and highest-frequency normal modes. Typical normal modes in the frequency bands, labeled with numbers in Fig. \ref{dos_part}, are visualized with their eigenvectors $\bm{a}_j^k/\sqrt{m_j}$ in Fig. \ref{map_eigenvec}. Small number of particles at the lowest and highest frequencies has high eigenvectors, indicating the localization of these modes. These localized normal modes are much different from the normal modes found in the crystals, i.e., phonons, and seems to be caused by the amorphous structure of the system. Similar low-frequency localized modes were also observed by the recent simulation \cite{widmer_2008} and experiment \cite{tan_2012}.

\section{Acoustic wave propagation}
Here, we investigated the acoustic wave propagation. The dispersion relation can be found from the dynamic structure factor; however, it can also be calculated by the normal mode analysis using the spectral density $S(k,\omega)$ which is a quantity proportional to the dynamic structure factor. The method of Taraskin et al. \cite{taraskin_2000} was employed in the present study. If an impinging plane wave with wavenumber $k$ is assumed in Eq. (\ref{input_w}), the contribution of mode $j$ to that plane wave can be applied in Eq. (\ref{alpha_kj}) as a spectral density coefficient $\alpha_{\bm{k}}^j$. The term $A$ in Eq. (\ref{input_w}) is a normalization coefficient, and a longitudinal wave and a transverse wave can be distinguished by whether the polarization vector $\hat{\bm{n}}$ is parallel or perpendicular to the wavenumber vector $\bm{k}$, respectively. Then, the coefficient $\alpha_{\bm{k}}^j$ provides the spectral density $S(k,\omega)$, which is calculated using Eq. (\ref{Skw}) \cite{taraskin_2000}. Here, we determined the coefficient $A$ from the normalization condition $\sum_{j=1}^{3N}|\alpha_{\bm{k}}^j|^2 = 1$.
\begin{equation}
\bm{w}_{\bm{k},i}=A\hat{\bm{n}}\cos(\bm{k}\cdot\bm{r}_i) 
\label{input_w}
\end{equation}
\begin{equation}
\alpha_{\bm{k}}^j=\sum_{i}^{N}\sqrt{m_i}\bm{a}_i^j \cdot \bm{w}_{\bm{k},i} 
\label{alpha_kj}
\end{equation}
\begin{equation}
S(k,\omega)=\frac{1}{3N}\sum_j^{3N}\left|\alpha_{\bm{k}}^j\right|^2\delta(\omega-\omega_j)
\label{Skw}
\end{equation}

\begin{figure*}[t]
\begin{center}
\subfigure[]{\includegraphics*[scale=0.62]{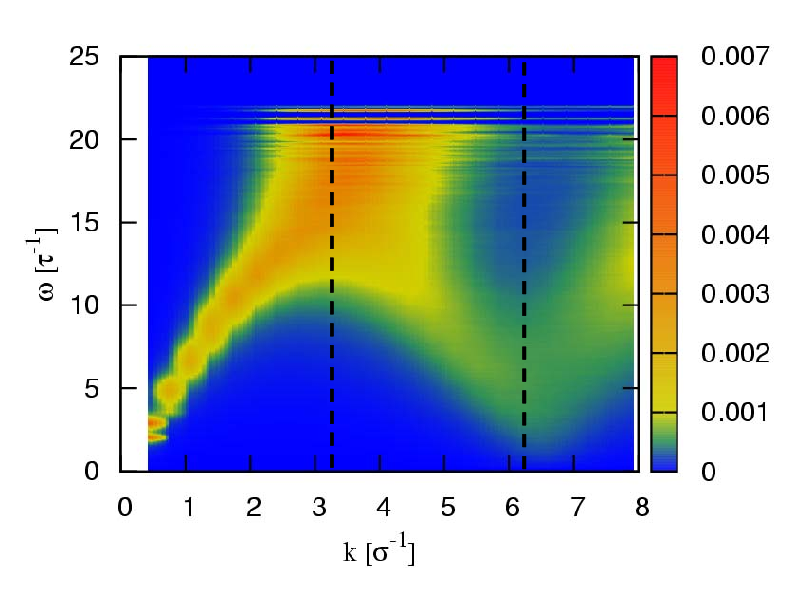}
}%
\subfigure[]{\includegraphics*[scale=0.62]{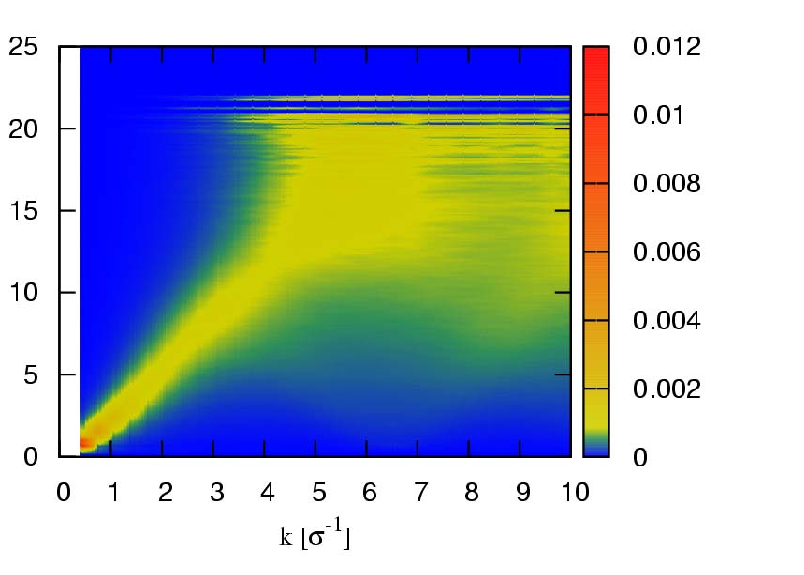}
}\hspace{-120cm}
\end{center}
\vspace{-0.25cm}
\caption{
Wavenumber and frequency dependences of the spectral density $S(k,\omega)$. Intense red color indicates higher values. (a) Excitation by longitudinal waves. Dashed lines represent $k=Q_p \approx 2\pi$ and $k=Q_p/2 \approx \pi$. (b) Excitation by transverse waves.
}
\label{pic_wS}
\end{figure*}

Figure \ref{pic_wS} shows the dependences of the spectral density on the wavenumber $k$ and the frequency $\omega$; it corresponds to a two-dimensional plot of the dynamic structure factor. Deeper red color indicates a greater value. The wavenumber and frequency dependences indicated in this figure are qualitatively very similar to those of amorphous glasses found in previous studies \cite{mazzacurati_1996,ruocco_2000}. The wavenumber $k=Q_p\approx 2\pi$ at which the static structure factor shows its first peak and the half-wavenumber $k=Q_p/2\approx \pi$ are indicated by the dashed lines in Fig. \ref{pic_wS}(a) for longitudinal waves. The region between $k=0$ and $Q_p/2$ is called the first (pseudo-)Brillouin zone and is an essential wavenumber region for investigating phonon-related phenomena \cite{grest_1984}.
Here we mention that several studies \cite{lewis_1994,muranaka_1994,hiwatari_1995,horbach_1996} investigated the finite size effects in terms of the sound wave propagation.
They found the artificial oscillations of the density correlation functions in $ortho-$terphenyl \cite{lewis_1994}, amorphous $\text{SiO}_2$ \cite{horbach_1996}, and 2-dimensional soft-sphere liquid \cite{muranaka_1994,hiwatari_1995}.
It was explained that such the artificial oscillation can be caused by the sound wave which propagates through the system and reappears due to the periodic boundary condition.
Unlike these studies \cite{lewis_1994,muranaka_1994,hiwatari_1995,horbach_1996}, we observed no artificial oscillations in our system (3-dimensional soft-sphere liquid).
We also confirmed that the sound wave attenuates completely within the system size $L$ and does not reappear through the periodic boundary.
Therefore, we consider that our result includes no finite size effects and catches the physically correct wave propagation.

\begin{figure*}[t]
\begin{center}
\subfigure[]{\includegraphics*[scale=0.41,angle=-90]{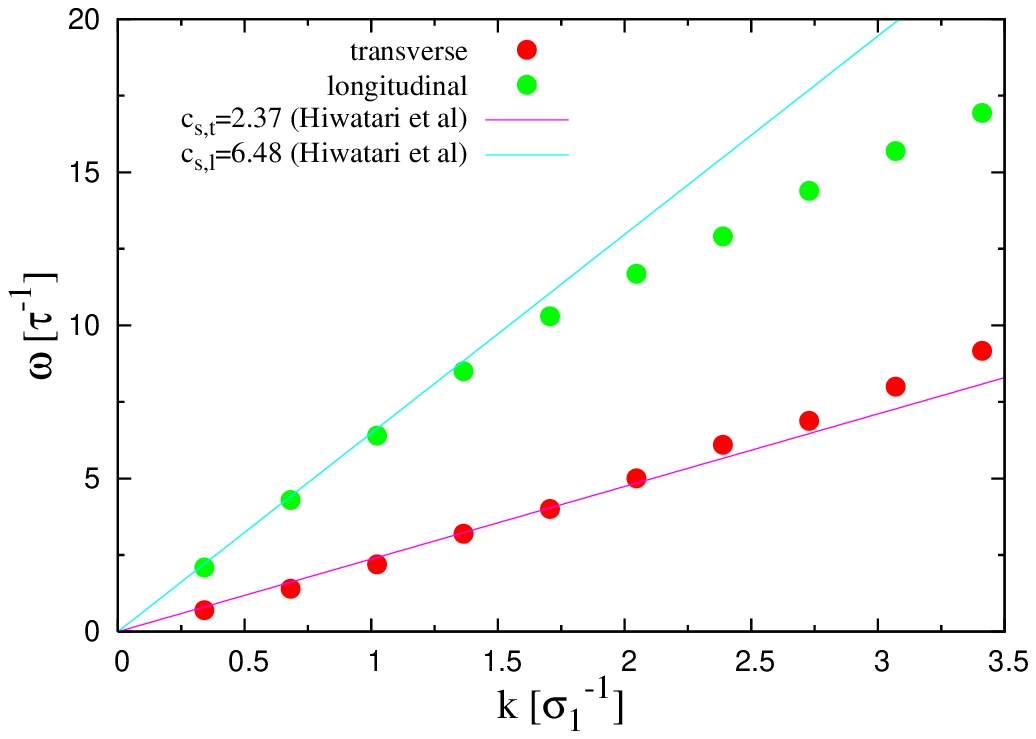}
\label{phonon_disp}
}
\subfigure[]{\includegraphics*[scale=0.41,angle=-90]{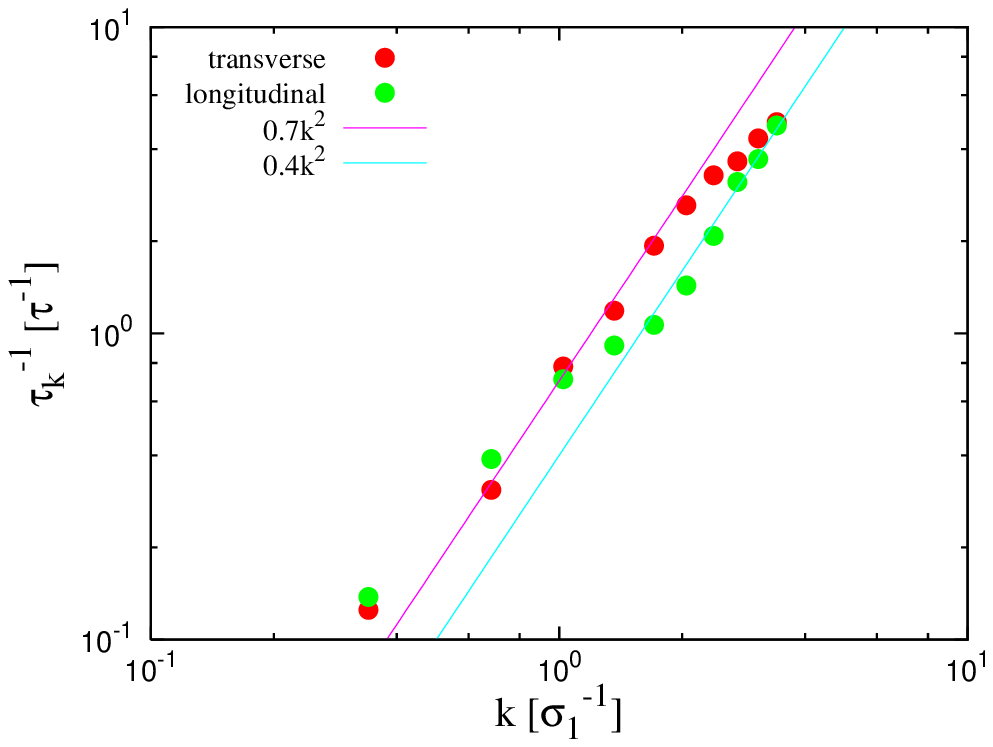}
\label{tau_k}
}\hspace{-100cm}
\end{center}
\vspace{-0.25cm}
\caption{
(a) Dispersion relation under initial excitation of the first Brillouin zone by both longitudinal and transverse waves in a planar wave. The solid line denotes a straight line based on the speed of sound found from the dynamic structure factor obtained by Hiwatari et al. \cite{hiwatari_1990} (b) Wavenumber dependence (double log scale) of reciprocal of attenuation time constant, $\tau_k^{-1}$. Solid line indicates the second power of the wavenumber, $\tau_k^{-1} \sim k^2$, for reference.
}
\end{figure*}

Figure \ref{phonon_disp} shows the dispersion relation determined from the wavenumbers at the peak values of the spectral density $S(k,\omega)$. This figure shows that the dispersion relation is well approximated by a linear relationship $\omega=c_sk$ in the first Brillouin zone, where the wavenumber is low and the wavelength is long. This linear relationship is characteristic of the dispersion relation of phonons in a crystalline structure; thus, it is possible to determine the speed of sound $c_s$ from the gradient of this line. For comparison, a line representing the speed of sound found from the dynamic structure factor in the supercooled state of the same type of system in a previous study \cite{hiwatari_1990} is drawn in Fig. \ref{phonon_disp}. The relationship obtained in the present study (the inherent structure) is quite close to the latter line (the supercooled state). This result confirms that the acoustic wave propagation through supercooled liquids are mostly determined by the inherent structure. It is worth mentioning that the recent experiment \cite{monaco_2009} and simulation \cite{monaco2_2009} demonstrated the deviation from a linear dispersion relation at very low frequency region in glasses, where the phase velocity is smaller than the macroscopic one $c_s$. They related such the ``softening'' of the sound velocity to universal anomalies in the low-frequency vibrational density of states, i.e., the so-called boson peak \cite{monaco_2009,monaco2_2009}. In the present study, we did not find the softening region of the sound velocity. We consider that this is because our inherent structure was obtained from the supercooled state with relatively large inherent energy \cite{sastry_1998}, so the boson peak and the softening region associated with it are shifted away to the lower frequency regime \cite{grigera_2003}. In addition, according to Ref. \citenum{monaco2_2009}, we have to use much larger system and reach considerable low frequency regime in order to find out the softening region.

Figure \ref{tau_k} shows the wavenumber dependence of the attenuation time constant $\tau_k$ for the plane wave associated with the wavenumber $k$, which was calculated using the half-width $\Gamma(k)$ of the spectral density $S(k,\omega)$. This figure shows that the short-wavelength vibrations of the high wavenumber components are damped out within a short time. The time constant $\tau_k$ approximately obeys the second power law behavior $\tau_k^{-1}\sim k^2$ in the present frequency regime. It should be noted that several experimental studies found out the strong scattering behavior, i.e., the forth power law dependence $\tau_k^{-1}\sim k^4$ at the low frequency regime in glasses \cite{monaco_2009,ruffle_2003,ruffle_2006,masciovecchio_2006}. In addition, the most recent study \cite{monaco2_2009} revealed such the strong scattering by numerical simulation. Like the softening of the sound velocity, the strong scattering is considered to be related to the boson peak \cite{monaco_2009,monaco2_2009,ruffle_2003,ruffle_2006,masciovecchio_2006}. Again, it can be considered that as we used the inherent structure of the supercooled state, such the strong scattering regime is shifted away to the lower frequency regime \cite{grigera_2003}.

As suggested in Ref. \citenum{masciovecchio_2006}, there are two physical origins of the acoustic wave attenuation: one is the anharmonicity of the interparticle interactions in the low frequency regime, and the other is the structural disorder in the high frequency regime. In the normal mode analysis, we assume the linear harmonic approximation; therefore the sound attenuation should be due to the structural disorder, not due to the anharmonicity. In the disordered structure, an impinging plane wave can not be expressed by a single normal mode but rather by a superposition of different normal mode waves with different frequencies. It is important to emphasize again that in the case of crystals which have ordered structures, a plane wave is expressed by a single normal mode (phonon). By considering the acoustic wave propagation as a superposition of normal modes, we can give a clear picture on the mechanism of the sound attenuation. If a plane wave is introduced in an amorphous structure at $t=0$, it is immediately decomposed into several different normal modes distributed within broad spectrum of frequencies for $t>0$. This means that the plane wave can not support its initial shape and begins to attenuate. Therefore, we can consider the sound attenuation as a phenomenon that one acoustic wave is decomposed to several different normal modes. As our results show the second power law dependence of the sound attenuation, $\tau_k^{-1} \sim k^2$, in Fig. \ref{tau_k}, we can associate this send power law dependence with the structural disorder origin \cite{ruocco_2000,horbach_2001}. We note that in the low frequency regime, the linear harmonic approximation can break down, and the anharmonicity can appear and contribute to the sound attenuation \cite{masciovecchio_2006}.

We remark that the dispersion relation resembling those for crystals was obtained for long-wavelength and low-frequency vibrations in Fig. \ref{phonon_disp}; they mimic the behavior of phonons in crystals. However, as demonstrated in Fig. \ref{map_eigenvec}, the amorphous structure of this system generates highly localized low-frequency normal modes. Therefore, as the acoustic wave consists of such localized normal modes, it is much different from phonon even if the dispersion relation is similar as that of phonon. Furthermore, the eigen frequencies of these low-frequency normal modes are less than the minimum frequencies (longitudinal waves: $2.10$; transverse waves: $0.698$) of phonons corresponding to the minimum wavenumber ($k_{min}=2\pi/L=0.342$) determined by the size $L$ of the system, which is found from the linear relationship between the frequency and the wavenumber. These lie outside the range of phonons. In the disordered system, these types of low-frequency localized modes contribute to the acoustic wave propagation.

\section{Concluding remarks}
In the present study, we investigated the acoustic wave propagation through a supercooled model liquid by means of normal mode analysis on the inherent structure. In the case of perfect crystals which have periodic and ordered structures, an acoustic wave is described by a phonon which is represented by a single normal mode of the crystals, therefore the acoustic wave propagate without attenuation. On the contrary, in the disordered materials, an acoustic wave decomposed into several different normal modes with different frequencies, which is quantified as the spectral density $S(k,\omega)$ in Eq. (\ref{Skw}) \cite{taraskin_2000}. The acoustic wave attenuates quickly in supercooled liquids, and this rapid attenuation is caused by the linear coupling of amorphous normal modes, originated by their disordered inherent structures, with a propagating plane wave. This also results in anomalous wavenumber dependency of the dispersion relation and the rate of attenuation. We will discuss further on this point more in detail by using a recently developed constitutive equation \cite{mizuno2_2011}.

At low frequencies, the acoustic wave consists of normal modes with relatively narrow frequency range, so it can propagate over some notable distances while it finally attenuates. Therefore, not much difference seems to exist between acoustic waves in supercooled liquids and crystals at low frequencies at macroscopic level. However, it is worth noting that the low-frequency acoustic wave is already composed of some highly localized normal modes as shown in Fig.2(a). This is of course very much different from the usual phonons in crystals. So far we do not yet understand well the role of the localized modes in low-frequency acoustic wave in supercooled liquids, we would like to emphasize that some notable differences appear between acoustic waves in supercooled liquids and crystals at mesoscopic and microscopic level.

\begin{acknowledgment}
This work was supported by KAKENHI 23244087 and the JSPS Core-to-Core Program, ``International research network for non-equilibrium dynamics of soft matter".
\end{acknowledgment}

\bibliographystyle{jpsj}
\bibliography{thesis}


\end{document}